\documentclass[11pt]{article}
\usepackage{amsmath}
\usepackage{amsfonts}

\usepackage{amssymb,graphicx}

\usepackage{amssymb}

\newcommand{\koniec}{\begin{flushright}  $\Box $ \end{flushright}}
\def\be{\begin{equation}}
\def\ee{\end{equation}}

\begin{document}

\title{A note on Penrose limits}
\author{Paul Tod\footnote{email: tod@maths.ox.ac.uk }\\Mathematical Institute,\\Oxford University}
%

\maketitle

\begin{abstract}
As a footnote to ref \cite{t1}, I show that, given a (time-like) umbilic 3-surface $\Sigma$ in a 4-dimensional space-time $M$, the Penrose limit taken along any null geodesic $\Gamma$ which lies in $\Sigma$ is a diagonalisable plane-wave.

\end{abstract}
\section{Introduction}
Given a space-time $M$ and a null geodesic $\Gamma$ in $M$, Penrose \cite{p1} defined a limiting process which produces a plane-wave space-time. One can ask whether the various plane-wave limits of an $M$ are \emph{diagonalisable}, which is to say whether coordinates can be found in which the metric has only diagonal entries. In 4-dimensions a simple algorithm for deciding whether a particular geodesic $\Gamma$ leads to a diagonalisable plane-wave was given in \cite{t2}, and in \cite{t1} I considered the problem of finding all 4-dimensional space-times with the property that \emph{all} of their  plane-wave limits were diagonalisable. In the process of finding examples to give in \cite{t1}, I realised that the following Proposition must hold:

\medskip

\noindent {\bf{Proposition}} \emph{Given a (time-like) umbilic 3-surface $\Sigma$ in a 4-dimensional space-time $M$, the Penrose limit taken along any null geodesic $\Gamma$ which lies in $\Sigma$ is diagonalisable.}

\medskip

Recall that a hypersurface $\Sigma$ is \emph{umbilic} if its second fundamental form is proportional to its intrinsic metric. An umbilic hypersurface is \emph{totally-geodesic} for null geodesics: any null geodesic from a point $p\in\Sigma$  and initially tangent to $\Sigma$ remains in $\Sigma$. A stronger condition on $\Sigma$ is for it to be \emph{extrinsically flat}, that is the second fundamental form is zero and such a $\Sigma$ is totally-geodesic for any geodesic. Evidently a totally-geodesic surface is umbilic. One could obtain an umbilic surface as a surface orthogonal to a hypersurface-orthogonal conformal Killing vector, for example a dilatation in flat space, and a totally-geodesic surface as the fixed point set of an involution, for example $\theta\rightarrow\pi-\theta$ in a spherically-symmetric metric when the equatorial plane is totally-geodesic, or as a hypersurface orthogonal to a hypersurface-orthogonal Killing vector, for example the surfaces of constant $x,y$ or $z$ in the Kasner metric:
\[g=dt^2-t^{2p}dx^2-t^{2q}dy^2-t^{2r}dz^2.\]
Thus this Proposition explains two results from \cite{t1}: why all Penrose limits of the Schwarzschild metric are diagonalisable - since any geodesic in Schwarzschild may be supposed w.l.o.g. to lie in the equatorial plane - and why the Penrose limit of the Kasner metric for a $\Gamma$ orthogonal to one of the Killing vectors is diagonalisable. 

\medskip

To prove the Proposition I'll first review how to take the Penrose limit, and then consider umbilic surfaces.

\medskip

\noindent{\bf{Acknowledgement:}} I realised the truth of this Proposition following a talk I gave at the 7th POTOR in \L\'od\'z in September 2021 and I am grateful to the organisers for the invitation and hospitality.

\section{The Penrose limit and umbilic surfaces.}
The 4-dimensional plane-wave metric can be written in the Brinkman form as
\be\label{1}
g=2du(dv+H(u,\zeta,\overline{\zeta})du)-2d\zeta d\overline{\zeta},\ee
with
\be\label{2}
H(u,\zeta,\overline{\zeta}))=\frac12\left(\Psi(u)\zeta^2+2\Phi(u)\zeta\overline{\zeta}+\overline{\Psi}(u)\overline{\zeta}^2\right).\ee
To take the Penrose limit of a space-time $M$ given a null geodesic $\Gamma$ in $M$, one first chooses a null vector $\ell^a$ tangent to $\Gamma$ and factorising into spinors as $\ell^a=\alpha^A\overline{\alpha}^{A'}$ and then rescales $\alpha^A$ so as to be parallelly-propagated along $\Gamma$ in the sense that
\[D\alpha^A:=\ell^{BB'}\nabla_{BB'}\alpha^A=0,\]
and an affine parameter $u$ along $\Gamma$ with $Du=1$. Then one calculates the two curvature components
\[\Psi(u)=\Psi_{ABCD}\alpha^A\alpha^B\alpha^C\alpha^D,\;\;\Phi(u)=\Phi_{ABA'B'}\alpha^A\alpha^B\overline{\alpha}^{A'}\overline{\alpha}^{B'},\]
where $\Psi_{ABCD},\Phi_{ABA'B'}$ are respectively the Weyl and Ricci spinors of $M$. Finally one substitutes the $\Psi$ and $\Phi$ thus obtained into $H$ in (\ref{2}) and then into the plane-wave metric (\ref{1}).

\medskip

It was shown in \cite{t2} that a plane-wave metric is diagonalisable if and only if the phase of $\Psi$ is constant along $\Gamma$. This can be equivalently stated as iff $\Psi(u)$ is real after a suitable constant rescaling of $\alpha^A$.

\medskip

Now suppose $\Sigma$ is a hypersurface with unit normal $N^a$ which we'll take to be space-like (i.e. $ N^aN_a=-1$ with our conventions)  so that the metric of $\Sigma$ is indefinite and there are null vectors tangent to $\Sigma$ (and therefore null geodesics of the ambient space $M$ lying entirely in $\Sigma$). The covariant derivative of $N_a$ can be written
\[\nabla_aN_b=K_{ab}-N_aA_b,\]
where $K_{ab}$ is the second fundamental form of $\Sigma$ and $A_b$ is the acceleration of the normals. Since $\Sigma$ is umbilic,
\[K_{ab}=\frac13Kh_{ab}\]
where $h_{ab}=g_{ab}+N_aN_b$ is the metric intrinsic to $\Sigma$ and $K=h^{ab}K_{ab}$.

\noindent Introduce the projection $P_a^{\;b}=h_a^{\;b}=\delta_a^{\;b}+N_aN^b$ orthogonal to $N^a$ and project the Ricci identity in the form
\[(\nabla_a\nabla_b-\nabla_b\nabla_a)N_c=R_{abcd}N^d\]
where $R_{abcd}$ is the Riemann tensor, orthogonal to $N$ on all indices. We obtain
\[    \frac13 P_a^{\;p}P_b^{\;q}P_c^{\;r}(K_{,p}h_{qr}-K_{,q}h_{pr})=        P_a^{\;p}P_b^{\;q}P_c^{\;r}N^dR_{pqrd},\]
and the trace-free part of this gives
\be\label{h1}P_a^{\;p}P_b^{\;q}P_c^{\;r}N^dC_{pqrd}=0,\ee
where $C_{pqrd}$ is the Weyl tensor. Recall the definition of the electric and magnetic parts of the Weyl tensor at $\Sigma$ as respectively
\[E_{ab}=C_{acbd}N^cN^d,\;\;H_{ab}=\frac12\epsilon_{ac}^{\;\;\;\;pq}C_{pqbd}N^cN^d,\]
then in particular (\ref{h1}) means that $H_{ab}$, the magnetic part of the Weyl tensor, is zero at $\Sigma$. (For the totally-geodesic case, we could have deduced the vanishing of $H_{ab}$ from the fact that 
$\epsilon_{abcd}$ changes sign under reflection).

\medskip

Next we choose a null geodesic $\Gamma$ that lies in $\Sigma$. Suppose its tangent is $\ell^a=\alpha^A\overline{\alpha}^{A'}$ with $D\alpha^A=0$ as before. Since $\Gamma$ lies in $\Sigma$ we'll have $N_a\ell^a=0$ and so
\[\alpha^AN_{AA'}=f\overline{\alpha}_{A'}\]
for some (nonzero) $f$. Taking $D$ of this equation and using $K_{ab}=\frac13Kh_{ab}$ we find $Df=0$ so $f$ is constant along $\Gamma$, and then a constant phase change of $\alpha^A$ will make $f$ real. Now we consider the identity
\[E_{ab}-iH_{ab}=2\Psi_{ACBD}\epsilon_{A'C'}\epsilon_{B'D'}N^{CC'}N^{DD'},\]
where $\Psi_{ABCD}$ is the Weyl spinor of $M$, and impose the vanishing of $H_{ab}$ on this, then contract with $\ell^a\ell^b=\alpha^A\overline{\alpha}^{A'}\alpha^B\overline{\alpha}^{B'}$ to find
\[E_{ab}\ell^a\ell^b=2f^2\Psi_{ABCD}\alpha^A\alpha^B\alpha^C\alpha^D=f^2\Psi(u).\]
To prove the Proposition, we need to establish that $\Psi(u)$ is real (possibly after a constant phase change on $\alpha^A$) and that now follows since both $f$ and $E_{ab}\ell^a\ell^b$ are real. Thus any $\Gamma$ lying in $\Sigma$ has a real $\Psi(u)$ after suitable choice of spinor $\alpha^A$, and therefore has a diagonalisable Penrose limit.
\koniec

\end{document}